# Exciton dynamics, Fano quantum interference and *d-d* excitation in the single crystal of two-dimensional antiferromagnetic Fe$_2$P$_2$S$_6$

Nasaru Khan[1,*], Yuliia Shemerliuk[2], Sebastian Selter[2], Bernd Büchner[2,3], Saicharan Aswartham[2] and Pradeep Kumar[1,†]

[1] *School of Physical Sciences, Indian Institute of Technology Mandi, Mandi-175005, India*
[2] *Leibniz-Institute for Solid-state and Materials Research, IFW-Dresden, 01069 Dresden, Germany*
[3] *Institute of Solid State and Materials Physics and Würzburg-Dresden Cluster of Excellence ct.qmat, Technische Universität Dresden, 01062 Dresden, Germany*

**Abstract**

Excitonic quasiparticle and their interactions with phonons, magnons and charge carriers may play a pivotal role in governing the optical properties and their correlation with magnetic interactions in two-dimensional (2D) magnetic semiconductors. Further, in transition metal compounds, *d-d* electronic transitions, arising from excitations between crystal-field split *d*-orbitals, significantly influence the optical and magnetic properties, particularly in strongly correlated and low-dimensional systems. Fe$_2$P$_2$S6, a layered antiferromagnetic semiconductor, offers a rich platform for studying the interplay between spin, charge, and lattice degrees of freedom in these 2D systems. In this work, we investigate the photoluminescence (PL) properties of Fe$_2$P$_2$S$_6$ to probe the exciton dynamics, intra-atomic transitions, and their temperature evolution. Two prominent *d-d* emission peaks are observed at ~ 1.63 eV (*D1*) and ~ 1.80 eV (*D2*), attributed to the crystal field-split Fe$^{2+}$ states. An excitonic emission near the band edge is also identified, which exhibits a characteristic Fano asymmetric line shape. This asymmetry is attributed to the quantum interference between the discrete excitonic state and the *d-d* transition induced continuum (*D2*), revealing a Fano resonance behaviour. This exciton peak disappears well before the Néel temperature, indicating its faster destabilization than magnetic ordering. Temperature-dependent PL measurements show a quenching of the excitonic peak with increasing temperature. Our findings provide detailed insight into the optical excitation pathways in Fe$_2$P$_2$S$_6$.

*nasarukhan736@gmail.com
†pkumar@iitmandi.ac.in

# 1. Introduction

Two-dimensional (2D) semiconducting materials have garnered substantial interest as viable alternatives to conventional silicon-based electronics, owing to their remarkable functional diversity. These materials offer distinct advantages such as scalable large-area synthesis, ultra-thin geometries, atomically smooth surfaces, mechanical cleavability into flexible layers, high charge carrier mobility, and energy band gap modulation governed by thickness. Among them, transition-metal dichalcogenides (TMDCs) with the general formula $MX_2$ (M = W, Mo, Re; X = S, Se) [1-5] have dominated research over the past decade. Their graphene-like layered structure, coupled with tunable electronic characteristics, strong spin-orbit coupling [6], valley-selective physics, and rich excitonic features such as excitons, trions, and biexcitons [7], make them strong contenders for next-generation electronic, optoelectronic, and chemical technologies.

In parallel, transition metal phosphorus trisulfides ($TM_2P_2S_6$), where TM = Ni, Fe, Mn, or Cu, represent a complementary family of 2D magnetic layered materials. These 2D magnetic materials have additional tuneable magnetic degrees of freedom, and positioning them as promising candidates for applications in electronics, spintronics, and optoelectronics. Among them, $Fe_2P_2S_6$ stands out having a band gap of ~ 1.5 eV and it crystallizes in the monoclinic phase [8]. Extensive studies have explored its potential across domains such as spintronics [9-10], catalysis [11], and device engineering [12]. Earlier studies have shown the persistence of antiferromagnetic ordering down to the monolayer limit in $Fe2P_2S_6$. This material features an Ising-like spin configuration with a Néel temperature ($T_N$) of ~ 120 K, which remains unaffected by layer thickness [13-14]. Within individual layers, $Fe^{2+}$ ions exhibit ferromagnetic interactions with two of their three nearest neighbours and antiferromagnetic coupling with the third, resulting in a zig-zag antiferromagnetic chains. These chains further couple antiferromagnetically within the plane and along the crystallographic *c*-axis, giving rise to a robust and intricate magnetic structure [15-17].

The understanding of excitonic and phononic behaviour in layered 2D magnetic semiconductors such as $Fe_2P_2S_6$ is driven by the need to unravel fundamental quasiparticle dynamics that underpin emergent phenomena in these 2D magnetic materials. While TMDCs have dominated excitonic research, the magnetic richness and structural complexity of $TM_2P_2S_6$ compounds remain comparatively unexplored. $Fe_2P_2S_6$, with its interplay of spin, orbital, and lattice degrees of freedom, presents a unique platform to investigate temperature-sensitive photoluminescence (PL) responses, exciton-phonon interactions, and spin-correlated electronic transitions. Understanding these processes is crucial not only for decoding intrinsic material behavior but also for guiding the design of future spin-optoelectronic devices. While extensive PL investigations have been conducted on $Ni_2P_2S_6$ within the $TM_2P_2S_6$ family, similar studies on $Fe_2P_2S_6$ are still lacking.

In this work, we present a comprehensive temperature-dependent PL study on bulk single crystals of $Fe_2P_2S_6$ to elucidate its intrinsic excitonic behaviour, exciton-phonon coupling, *d-d* transitions, and the temperature-induced PL quenching phenomenon. Temperature-dependent PL spectroscopy has proven to be a powerful technique for probing excitonic quasiparticle dynamics in 2D materials [18-23]. Understanding exciton relaxation mediated by phonon interactions is not only of fundamental significance but also critical for advancing optoelectronic device applications. Our findings also underscore the pronounced influence of acoustic phonons in governing exciton relaxation processes in $Fe_2P_2S_6$.

## 2. Experimental Details

PL measurements were performed using a LabRAM HR Evolution micro-Raman spectrometer in a backscattering configuration. The sample was excited with a 633 nm laser (~1.96 eV), focused using a 50× long working distance objective. To minimise the local heating, the laser power was kept below 0.5 mW. Inelastically scattered light was dispersed using a 600 grooves/mm grating and detected by a Peltier-cooled CCD. Temperature control was achieved over a range of 4-300 K with a precision of ±0.1 K using a closed-cycle refrigerator (Montana Instruments).

## 3. Results and discussion

### 3.1 Photoluminescence of bulk single crystals of Fe$_2$P$_2$S$_6$

Bulk Fe$_2$P$_2$S$_6$ crystallizes in a monoclinic structure, belonging to the point group $C_{2h}$ and space group $C2/m$, #12. The crystal structure is layered, exhibiting rich interplay between magnetic and structural motifs. Magnetic Fe atoms form a honeycomb network, each coordinated octahedrally by six sulphur atoms, with some finite trigonal distortion. Phosphorus atoms are bonded to three sulphur atoms and one additional phosphorus atom, forming a [P$_2$S$_6$]$^{4-}$ complex anion (see Figure 1 (a)). Figure 1(b) presents the PL spectra of bulk Fe$_2$P$_2$S$_6$ at various temperatures. At 4 K, seven distinct peaks emerge, labelled for convenience as P1-P4, D1-D2, and E1. The intensities of these peaks decrease progressively with increasing temperature. Peaks P1-P4 and E1 vanish at ~ 70 K, well below the Néel temperature (T$_N$ ~ 120 K), whereas the broader peaks D1 and D2 features persist up to room temperature. The low-energy peaks P1-P4 are attributed to the phonon sidebands associated with the excitonic emission. The dominant PL feature, E1, appears as an asymmetric peak centred at ~ 1.853 eV, and is identified as an excitonic emission. Its asymmetric line shape is understood invoking Fano resonance, which will be discussed in detail later. Interestingly, the energy of this excitonic peak exceeds the reported bandgap of Fe$_2$P$_2$S$_6$, indicating that this emission is not due to direct band-to-band recombination. This observation suggests that the luminescence may involve electronic states located above the conduction-band minimum.

## 3.2 Temperature dependence of the bandgap

To investigate the temperature dependence of the exciton dynamics in $Fe_2P_2S_6$, we recorded PL spectra across a broad range of temperature. These spectra were fitted using Lorentzian line shape function except excitonic peak E1, which is fitted using Bright-Wigner Fano (BWF) function (see details in section 3.5). Figure 2(a) illustrates the temperature evolution of the excitonic peak E1, where a clear redshift in peak energy is observed with increasing temperature, reflecting a characteristic behaviour typical of conventional semiconductors. The temperature-induced variation in bandgap energy primarily arises from two mechanisms: (1) Relative shift of the conduction and valence band edges due to lattice dilation with increasing temperature [24]. This mechanism leads to a linear decrease in the bandgap at higher temperatures and nonlinear variations in the lower temperature range. (2) Electron-phonon interactions, which induce band-edge renormalization, generally resulting in a reduction of bandgap energy with increasing temperature [25-26]. In $Fe_2P_2S_6$, the crystal undergoes anisotropic thermal contraction below the Néel temperature i.e. the *b*-axis elongates, while *a* and *c* axes contract, culminating in an overall volume reduction [27]. This structural evolution may contribute to the observed decrease in bandgap with increasing temperature.

The temperature-dependent variation of the bandgap energy in semiconductors is often described by empirical models, one of which is the Varshni relation [28]. It is given by: $E(T) = E_0 - \frac{\sigma T^2}{T + \beta}$, where $E_0$ is the bandgap energy at 0 K, $\sigma$ quantifies exciton-phonon coupling strength, and $\beta$ denotes the Debye temperature of the material. The red solid line in Figure 2(a) represents the Varshni fit to the experimental data, showing a very good agreement. From fitting the values of $E_0$, $\sigma$ and $\beta$ comes out to be 1.853 eV, 4.058×10$^{-5}$ eVK$^{-1}$ and 176.8 K, respectively. An alternative empirical relation used to describe the temperature-dependent bandgap behaviour is the O'Donnell-Chen model [29], expressed as:

$E(T) = E_0 - S \times E_P[\coth(\frac{E_P}{2k_bT}) - 1]$. Here, $k_b$ is the Boltzmann constant, $E_0$ denotes the exciton peak energy at 0 K, $E_p$ represents the average phonon energy contributing to the bandgap renormalization; and $S$ is the dimensionless Huang-Rhys factor, which characterizes the strength of exciton-phonon coupling. The solid purple line in Figure 2(a) corresponds to the O'Donnell-Chen fit, demonstrating a very good agreement. Extracted values of the fitting parameters $E_0$, $S$ and $E_p$ are 1.853 eV, 0.09541 and 2.96 meV, respectively.

Figure 2(d) presents the temperature dependence of the integrated PL intensity for the excitonic peak E1. A marked decline in intensity is observed with increasing temperature, indicative of thermal quenching of excitonic emission. The temperature-dependent intensity behaviour is modelled using the relation: $I = \frac{I_0}{1 + Ae^{-E_B/k_BT}}$, where, $I_0$ and $E_B$ represent the intensity at 0 K and exciton binding energy, respectively. $k_B$, is the Boltzmann constant. The experimental data fit well to this model, yielding an exciton binding energy of ~ 11.6 meV.

**3.3 Thermal broadening of the exciton**

The thermally induced variation in the FWHM of an excitonic peak reflects the evolution of spectral broadening with temperature, and is intimately linked to both the exciton lifetime and its interactions with lattice vibrations and other quasiparticle excitations. Excitonic linewidth is influenced primarily by scattering processes involving longitudinal optical (LO) phonons, acoustic phonons, and thermal dissociation into free carriers. Additional contributions may arise from structural imperfections and impurity scattering. Figure 2(b) illustrates the temperature dependence of the linewidth of the excitonic peak E1. A pronounced homogeneous broadening is observed with increasing temperature, underscoring the role of exciton-phonon coupling.

At finite temperatures, the linewidth of an exciton can be modelled as [30]:

$$\Gamma(T) = \Gamma_0 + \Gamma_{AC}(T) + \Gamma_{LO}(T) = \Gamma_0 + \lambda_{AC}T + \lambda_{LO}N_{LO}(T)$$

Where $\Gamma_0$ denotes the intrinsic linewidth at 0 K, primarily due to impurity and defect scattering. $\lambda_{AC}$ and $\lambda_{LO}$ are coupling coefficients representing exciton-acoustic and exciton-LO phonon interactions, respectively. The LO phonon interaction is typically described by the Fröhlich mechanism [31], while acoustic phonon contributions are governed by the deformation potential model [30]. $N_{LO}(T)$ is the Bose-Einstein distribution function, given as $N_{LO}(T) = 1/[e^{E_{LO}/k_BT} - 1]$; where $E_{LO}$ representing the energy of the participating *LO* phonons. The experimental data is fitted well using this model, shown by the solid blue line in Figure 2(b). To disentangle the contributions of individual phonon interactions, Figure 3 presents the separate temperature-dependent effects of acoustic phonons (black spheres) and LO phonons (red square) on linewidth broadening. While both phonon types are involved, acoustic phonons dominate in the low-temperature regime (up to ~ 40 K), due to their lower energy and higher thermal population. Conversely, LO phonons-whose energy exceeds thermal energy at lower temperatures become more populated and increasingly influential in linewidth broadening above ~ 40 K. As a result, both acoustic and LO phonon interactions contribute to the spectral evolution of the excitonic peak.

**3.4 Exciton-phonon coupling**

Figure 4 displays the Lorentzian-fitted PL spectrum of $Fe_2P_2S_6$ recorded at 4 K. Notably, a series of satellite peaks, phonon sidebands, accompanying the main excitonic emission are observed. These sidebands are characteristic signatures of strong exciton-phonon coupling. This phenomenon aligns with the Franck-Condon principle, which posits that electronic transitions occur on timescales much faster than nuclear motion-leading to vertical transitions on the potential energy surface. As a result, electronic excitations are accompanied by phonon-assisted transitions, leading to a series of sidebands equally spaced by the energy of the coupled phonon mode. To quantify the energy separation between these peaks, the PL spectrum was fitted using Lorentzian line shapes. Our analysis revealed that these sidebands (P1-P4) are separated by a consistent energy difference of ~ 254 cm$^{-1}$ (~ 0.0315 eV). This shift closely matches with the phonon energy of the $A_g$ optical phonon mode observed in the Raman spectrum at ~ 248 cm$^{-1}$, affirming the participation of phonons in the exciton recombination process. The thermal quenching of these sidebands, concurrent with the disappearance of the excitonic peak at elevated temperatures, also supports their origin as phonon sidebands replicas rather than independent electronic transitions. Our observations underscore the pronounced exciton-phonon interaction in $Fe_2P_2S_6$.

We note that, exciton-phonon interactions have been extensively studied in various low-dimensional systems [32-33], revealing the emergence of phonon sidebands as a consequence of coupling between excitons and lattice vibrations. These sidebands often reflect the formation of excitonic-polaron states, wherein localized excitons are coupled to LO phonons. Taken together, these insights point to a compelling interplay between electronic and vibrational degrees of freedom in $Fe_2P_2S_6$, which fundamentally shapes its optical and magneto-optical properties.

### 3.5 *d-d* electronic transition in $Fe_2P_2S_6$

In crystalline solids, excitations between localized *3d* orbitals of transition metal ions, commonly referred to as *d-d* transitions, play a fundamental role in shaping a wide array of phenomena in condensed matter physics. In an isolated transition metal atom with an open *d* shell, the *3d* levels are five-fold degenerate, each associated with a distinct quadrupolar charge distribution in real space. In $Fe_2P_2S_6$, both the valence band maximum and conduction band minimum are predominantly composed of Fe *3d* states, enabling its classification as a Mott-Hubbard-type insulator [34-35]. Consequently, the low-energy electronic behaviour of this system is primarily governed by Fe $3d$ configurations. Under the influence of an octahedral crystal field in bulk $Fe_2P_2S_6$, the fivefold degenerate $d$ orbitals split into a lower-energy triplet $t_{2g}$ ( $d_{xy}$, $d_{yz}$, $d_{xz}$ ) and a higher-energy doublet $e_g$ ( $d_{x^2-y^2}$, $d_{z^2}$ ), see Fig. 5. Given that the $(P_2S_6)^{-4}$ ligand is weak in field strength, the crystal field splitting follows a high-spin configuration. As a result, within the split manifold, the higher-energy ($e_g$) orbitals are half-filled, while one of the ($t_{2g}$) orbitals is fully occupied and the remaining two are half-filled, leading to four half-filled orbitals per Fe site and a spin state of (S = 2). Notably, the Fe-S octahedra exhibit a slight trigonal distortion, lowering the local point group symmetry from $O_h$ to $D_{3d}$. This distortion introduces additional splitting of the $d$ orbitals, with an energy scale of ~ 0.35 eV [36]. Specifically, the crystal field elongates or compresses along the [111] direction [37-38], further lifting the degeneracy of the $t_{2g}$ states while retaining degeneracy among the $e_g$ states [37]. The $t_{2g}$ manifold further splits into an $a_{1g}$ singlet and an $e_{g\pm}^{\pi}$ doublet, which are linear combinations of the $d_{xy}$, $d_{yz}$, $d_{xz}$ orbitals given as $a_{1g} = \frac{1}{\sqrt{3}}(|xy\rangle + |yz\rangle + |xz\rangle)$

$e_{g\pm}^{\pi} = \frac{1}{\sqrt{3}}(|xy\rangle + \exp(\mp\frac{2i\pi}{3})|yz\rangle + \exp(\pm\frac{2i\pi}{3})|xz\rangle)$ [39].

The PL spectra of $Fe_2P_2S_6$ exhibit two broad features centred at ~ 1.63 eV (D1) and ~ 1.80 eV (D2). Unlike phonon sidebands, these peaks possess considerable FWHM and persist well above the Néel temperature (~ 120 K), remaining visible even at 300 K. Their resilience against thermal quenching and spectral broadening rules out attribution to magnetic continuum or spin dynamics. Instead, these features may be understood invoking $d$-$d$ electronic transitions. The only spin-allowed $d$-$d$ transition in $Fe_2P_2S_6$ corresponds to the excitation of an electron from a fully occupied $a_{1g}/e_{g\pm}^{\pi}$ to $e_g$ orbital. We assign the D1 peak at ~ 1.63 eV to this spin-allowed $d$-$d$ transition. Previous studies reported a crystal field splitting of only ~ 1.1 eV [40], while the additional trigonal distortion contributes ~ 0.35 eV, increasing the energy separation between the $a_{1g}/e_{g\pm}^{\pi}$ and $e_g$ orbital levels and making such transitions energetically plausible.

The D2 feature observed near ~1.80 eV is attributed to the spin-forbidden d–d transition, specifically a spin-flip process in which an electron transitions from a spin-up to a spin-down state within the same orbital manifold. This excitation generates a localized spin perturbation at the transition-metal site. The system may respond to this spin-flip excitation via two distinct pathways. In the first scenario, the excitation can propagate through the lattice via a virtual electron hopping mechanism pathway where the spin-flip perturbation is transferred to a neighboring site through an intermediate high-energy virtual state. Alternatively, the excitation may undergo local relaxation via spin-orbit coupling (SOC). Here, SOC mediates an interaction between spin and orbital degrees of freedom, facilitating the dissipation of spin excitation energy through a spin-flip event that restores the ground-state spin configuration. This process may be accompanied by orbital or lattice relaxation effects [40].

To analyse the temperature behaviour of peak D1, we fitted the peak using a Lorentzian line shape to extract self-energy parameters i.e. peak energy, FWHM, and the intensity. Figure 6 shows the temperature evolution of these parameters. The peak energy of D1 exhibits a non-monotonic trend i.e. first it increases from 4 K to ~150 K, followed by a gradual decrease up to 300 K. This behaviour suggests an intricate interplay between crystal field effects, spin-spin interactions, and electron-phonon coupling. Below $T_N$, Fe$_2$P$_2$S$_6$ order antiferromagnetically with a zigzag spin configuration within its 2D honeycomb layers [14, 41]. In this regime, high-spin Fe$^{2+}$ ions (3d$^6$ configuration) are subjected to both crystal field effects from the surrounding $(P^2S^6)^{-4}$ ligands and an effective exchange field arising from neighbouring Fe ions. The internal exchange field in the ordered phase may modify the $d$ orbital energies i.e. either further split the levels or reduce the energy of certain $d$-$d$ transitions due to spin alignment effects. Interestingly, the blueshift in the D1 peak persists beyond $T_N$, indicative of short-range spin-spin correlations common in quasi-2D magnetic systems [42-44]. These correlations continue to influence the electronic states and transition energies well above the long-range ordering temperature, sustaining the observed blueshift up to ~ 150 K.

Above ~ 150 K, AFM ordering is significantly suppressed due to thermal fluctuations, pushing the system into the paramagnetic regime. In this state, the internal exchange field no longer influences the energy of *d-d* excitations. Instead, the observed redshift in the D1 peak is primarily governed by lattice and phononic effects. Thermal expansion leads to elongation of Fe-S bonds, which weakens the octahedral crystal field strength and reduces the energy separation between *d*-orbitals, thereby lowering the energy of the spin-allowed *d-d* transition. Additionally, electron-phonon coupling may contribute to this spectral redshift by renormalizing the electronic states, further shifting the *d-d* excitation energy with increasing temperature [26, 28]. These plausible mechanisms may collectively account for the temperature-induced redshift of the D1 peak in the PL spectra.

**3.6 Fano quantum interference**

Quantum interference between a discrete state and a continuum (either magnetic or electronic) gives rise to the characteristic Fano asymmetric line shape, with the coupling strength governing the degree of spectral asymmetry. While Fano resonances have been widely observed using absorption, Raman, and transmission spectroscopy, their presence in PL spectra remains relatively rare. This phenomena of Fano resonances is quite ubiquitous and has been reported across diverse systems, including optomechanical resonators [45], semiconductor nanostructures [46-47], superconductors [48-49], photonic crystals [50-55], dielectric nanoparticles [56], and plasmonic nanoantennas [57-58].

The excitonic peak E1 in the PL spectrum exhibits a pronounced Fano-type asymmetric profile. This Fano resonance arises due to the coupling between the discrete excitonic state (E1) and the continuum-like D2 band originating from *d-d* transitions. The asymmetric profile of E1 was quantitatively analysed using the Breit-Wigner-Fano (BWF) function [59-61]:

$$I = I_0 \frac{[1+2(\omega-\omega_0)/(q\Gamma)]^2}{[1+4(\omega-\omega_0)^2/\Gamma^2]}$$

Where, $\omega_0$ is the uncoupled frequency, $\Gamma$ denotes the intrinsic linewidth, and (1/q) is the Fano asymmetry parameter. Fano asymmetry parameter characterizes the nature and strength of coupling large (|q|) (small (1/q)) indicates weak coupling and a near-symmetric (Lorentzian) peak, while small (|q|) (large (1/q)) signifies stronger coupling and pronounced asymmetry. Figure 2(c) presents the temperature evolution of the Fano asymmetry parameter (1/q), which decreases with increasing temperature. This trend reflects the weakening of the continuum (D2) due to thermal effects, resulting in diminished exciton-continuum coupling.

**Conclusion**

In summary, we carried out an in-depth photoluminescence study of the layered 2D antiferromagnetic semiconductor $Fe_2P_2S_6$, revealing rich excitonic and intra-atomic optical phenomena. Two well-defined *d-d* emission peaks at ~ 1.63 eV and ~ 1.80 eV, attributed to $Fe^{2+}$ crystal field transitions, were observed alongside a distinct excitonic emission near the band edge. Temperature-dependent PL measurements reveal that the excitonic emission is sensitive to thermal activation, while the *d-d* emission features remain robust, highlighting their localized character. The anomalous temperature evolution of peak energy of spin allowed *d-d* transition (*D1*) reflects a complex interplay between magnetic interactions, crystal field effects. Notably, the excitonic peak exhibits a Fano asymmetric line shape, indicating quantum interference between the discrete excitonic state and the continuum of the spin forbidden (*D2*) *d-d* transition. We observed that the phonon sideband appeared at the left of excitonic peak with equally spaced modes and disappears with excitonic peak, manifesting the exciton-phonon coupling. These results not only deepen our understanding of light-matter interaction in $Fe_2P_2S_6$ but also unravel the pivotal role of *d*-electron in controlling the properties in $Fe_2P_2S_6$, and highlight the role of quantum interference in shaping the excitonic response of 2D magnetic van der Waals semiconductors.

**Acknowledgement**

NK acknowledge CSIR India for the fellowship. PK acknowledge support from IIT Mandi for the experimental facilities and ANRF (CRG/2023/002069) for the financial support. SA acknowledges Deutsche Forschungsgemeinschaft (DFG) through Grant No. AS 523/4–1 and BB through SFB 1143 (project-id 247310070), ct.qmat (EXC 2147, project-id 390858490).

**References**

[1] M. Jhao, Y. Hao, C. Zhang, R. Zhai, B. Liu, W. Liu, C. Wang, S. H. M. Jafri, A. Razaq, R. Papadakis *et al.,* Crystals, **12**, 1087 (2022).

[2] F. M. Pinto, M. C. M. D. de Conti, W. S. Pereira *et al.*, Catalysts, **14**, 388 (2024).

[3] L. Yang, C. Xie, J. Jin, R. N. Ali, C. Feng, P. Liu, B. Xiang, Nanomaterials, **8**, 463 (2018).

[4] W. Choi, N. Choudhary, G. H. Han, J. Park, D. Akinwande, Y. H. Lee, Materials Today, **20**, 116 (2017).

[5] M. Fuhrer, J. Hone, Nature Nanotech**, 8**, 146 (2013).

[6] Z. Y. Zhu, Y. C. Cheng, and U. Schwingenschlögl, Phys. Rev. B, **84**, 153402 (2011).

[7] D. Kumar, R. Kumar, M. Kumar and P. Kumar, J. Mater. Chem. C, **10**, 5684 (2022).

[8] G. Ouvrard, R. Brec and J. Rouxel, Mater. Res. Bull., **20**, 1181 (1985).

[9] G. L. Flem, R. Brec, G. Ouvrard, A. Louisy, and P. Segransan, J. Phys. Chem. Solids, **43**, 455 (1982).

[10] K. C. Rule, G. J. McIntyre, S. J. Kennedy and T. J. Hicks, Phys. Rev. B, **76**, 134402 (2007).

[11] W. Zhu, W. Gan, Z. Muhammad, C. D. Wang, C. Q. Wu, H. J. Liu, D. B. Liu, K. Zhang, Q. He, H. L. Jiang, X. S. Zheng, Z. Sun, S. M. Chenand and L. Song, Chem. Commun., **54**, 4481 (2018).

[12] Y. Fujji *et al*., Electrochimica Acta, **243**, 370 (2017).

[13] J. U. Lee, S. Lee, J. H. Ryoo, S. Kang, T. Y. Kim, P. Kim, C. H. Park, J. Park, H. Gand Cheong, Nano Lett., **16**, 7433 (2016).

[14] P. A. Joy and S. Vasudevan, Phys. Rev. B, **46**, 5425 (1992).

[15] A. R. Wildes, K. C. Rule, R. I. Bewley, M. Enderle and T. J. Hicks, J. Phys.: Condens. Matter, **24**, 416004 (2012).

[16] M. Balkanski, M. Jouanne, G. Ouvrard and M. Scagliotti, J. Phys. C: Solid State Phys., **20**, 4397 (1987).

[17] B. L. Chittari, Y. Park, D. Lee, M. Han, A. H. MacDonald, E. Hwang and J. Jung, Phys. Rev. B, **94**, 184428 (2016).

[18] M. Tebyetekerwa, J. Zhang, Z. Xu, T. N. Truong, Z. Yin, Y. Lu, S. Ramakrishna, D. Macdonald, H. T. Nguyen, ACS Nano, **14**, 14579 (2020).

[19] O. D. P. Zamudio *et al*., 2D Mater., **2,** 035010 (2015).

[20] K. Wu, W. Sun, Y. Jiang, J. Chen, Li Li, C. Cao, S. Shi, X. Shen, J. Cui, J. Appl. Phys., **122**, 075701 (2017)

[21] T. Korn *et al.*, Appl. Phys. Lett., **99**, 102109, (2011).

[22] S. Tongay *et al.*, Nano Lett., **12**, 5576 (2012).

[23] S. Kang, et al., Nature, **583**, 785 (2020).

[24] J. Bardeen and W. Shockley, Phys. Rev., **80**, 72 (1950).

[25] H. Fan, Phys. Rev., **82**, 900 (1951).

[26] D. Olguın, M. Cardona, and A. Cantarero, Solid State Commun., **122**, 575 (2002).

[27] C. Murayama et al., J. Appl. Phys., **120**, 142114 (2016).

[28] Y. P. Varshni, Physica, **34**, 149 (1967).

[29] K. P. O'Donnell and X. Chen, App. Phys. Lett., **58**, 2924 (1991).

[30] S. Rudin *et al*., Phys. Rev. B: Condens. Matter Mater. Phys., **42**, 11218 (1990).

[31] R. M. Martin and T. C. Damen, Phys. Rev. Lett., **26**, 86 (1971).

[32] D. Christiansen *et al.,* Phys. Rev. Lett. **119**, 187402 (2017).

[33] S.-J. Xiong and S.-J. Xu, EPL, **71,** 459 (2005).

[34] J. E. Nitschke, D. L. Esteras, M. Gutnikov, K. Schiller, S. Mañas-Valero, E. Coronado, M. Stupar, G. Zamborlini, S. Ponzoni, J. J. Baldoví, M. Cinchetti, Mater. Today Electron., **6**, 100061 (2023).

[35] Y. Jin, M. Yan, T. Kremer, E. Voloshina and Y. Dedkov, Sci. Rep., **12**, 735 (2022).

[36] A. G. Chang, Phys. Rev. B, **106**, 125412 (2022).

[37] C. Ballhausen, Introduction to Ligand Field Theory, McGraw Hill Series in Advanced Chemistry (McGraw-Hill, New York, 1962).

[38] D. I. Khomskii, Transition Metal Compounds (Cambridge University Press, Cambridge, UK, 2014).

[39] B. Pal, Y. Cao, X. Liu, F. Wen, M. Kareev, A. T. N'Diaye, P. Shafer, E. Arenholz, and J. Chakhalian, Sci. Rep. **9**, 1896 (2019).

[40] J. E. Nitschke, Newton, **1,**100019 (2025).

[41] N. Khan et al., 2D Mater., **11**, 035018 (2024).

[42] N. Khan *et al.*, Sci Rep, **15**, 4438 (2025).

[43] J. Khatua, T. Arh, S. B. Mishra, *et al.*, *Sci Rep*, **11**, 6959 (2021).

[44] A.G. Chakkar *et al.*, Phys. Rev. B, **109**, 134406 (2024).

[45] K. Y. Fong, L. Fan, L. Jiang, X. Han, and H. X. Tang., Phys. Rev. A., **90**, 051801, (2014).

[46] C. Holfeld, F. Löser, M. Sudzius, K. Leo, D. Whittaker, and K. Köh ler., Phys. Rev. Lett., **81**, 874 (1998).

[47] P. Fan, Z. Yu, S. Fan and M. L. Brongersma., **13**, 471 (2014).

[48] M. Limonov, A. Rykov, S.Tajima, and A. Yamanaka., Phys. Rev. Lett., **80**, 825 (1998).

[49] V. Hadjiev, X. Zhou, T. Strohm, M. Cardona, Q. Lin, and C. Chu., Phys. Rev. B., **58**, 1043 (1998).

[50] A. E. Miroshnichenko, S. Flach and Y. S. Kivshar, Rev. Modern Phys., **82**, 2257 (2010).

[51] I. Soboleva, V. Moskalenko and A. Fedyanin., Phys. Rev. Lett., **108**, 123901 (2012).

[52] H. Yang, D. Zhao, S. Chuwongin, J.-H. Seo, W. Yang, Y. Shuai, J. Berggren, M. Hammar, Z. Ma, and W. Zhou, Nature Photonics., **6**, 615 (2012).

[53] M. Rybin, A. Khanikaev, M. Inoue, A. Samusev, M. Steel, G. Yushin, and M. Limonov, Photonics Nanostructures: Fun dam. Appl. **8**, 86 (2010).

[54] W. Zhou, D. Zhao, Y.-C. Shuai, H. Yang, S. Chuwongin, A. Chadha, J.-H. Seo, K. X. Wang, V. Liu, and Z. Ma, Prog. Quantum Electron., 38, 1 (2014).

[55] P. Markos, Phys. Rev. A, **92**, 043814 (2015).

[56] K. E. Chong, B. Hopkins, I. Staude, A. E. Miroshnichenko, J. Dom inguez, M. Decker, D. N. Neshev, I. Brener, and Y. S. Kivshar, Small, **10**, 1985 (2014).

[57] B. Hopkins, D. S. Filonov, S. B. Glybovski and A. E. Miroshnichenko., Phys. Rev. B., **92**, 045433 (2015).

[58] B. Hopkins, A. N. Poddubny, A. E. Miroshnichenko and Y. S. Kivshar., Laser Photonics Rev., **10**, 137 (2016).

[59] K. Kim, S. Y. Lim, J. U. Lee, S. Lee, T. Y. Kim, K. Park, G. S. Jeon, C. H. Park, J. Park, H. Gand Cheong, Nat. Commun., **10**, 335 (2019).

[60] U. Fano, Phys. Rev. B, **124,** 1866 (1961).

[61] P. H. Tan *et al.*, Nat. Mater., **11**, 294 (2012).

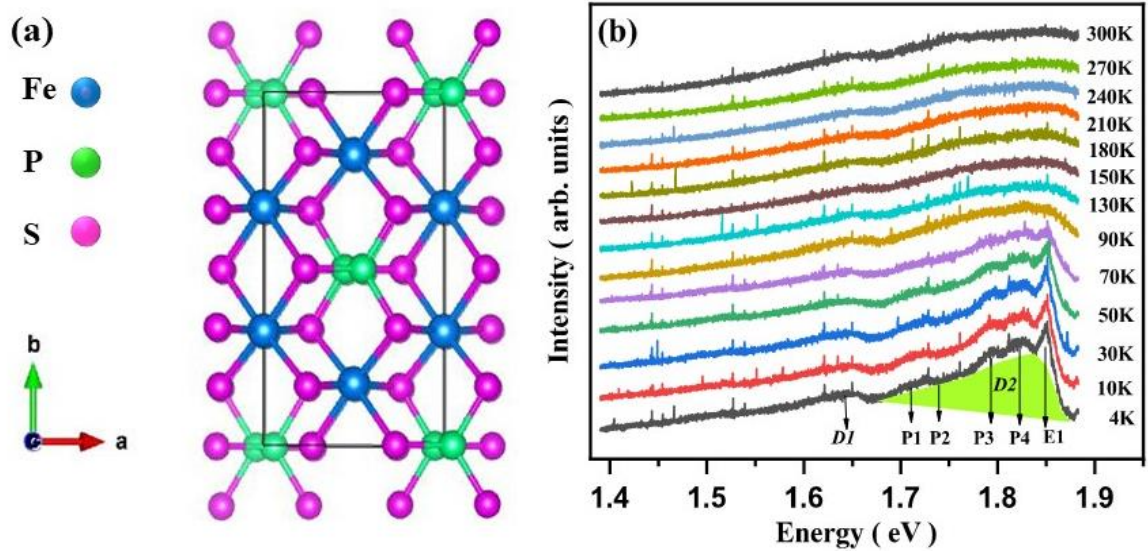

**Figure 1. (a)** Crystal structure of $Fe_2P_2S_6$ viewed along the *ab*-plane **(b)** Temperature-dependent PL spectra of bulk $Fe_2P_2S_6$. Distinct features include phonon sidebands (P1-P4), a pronounced excitonic peak (E1), and broad emission bands (D1 and D2) attributed to *d-d* electronic transitions.

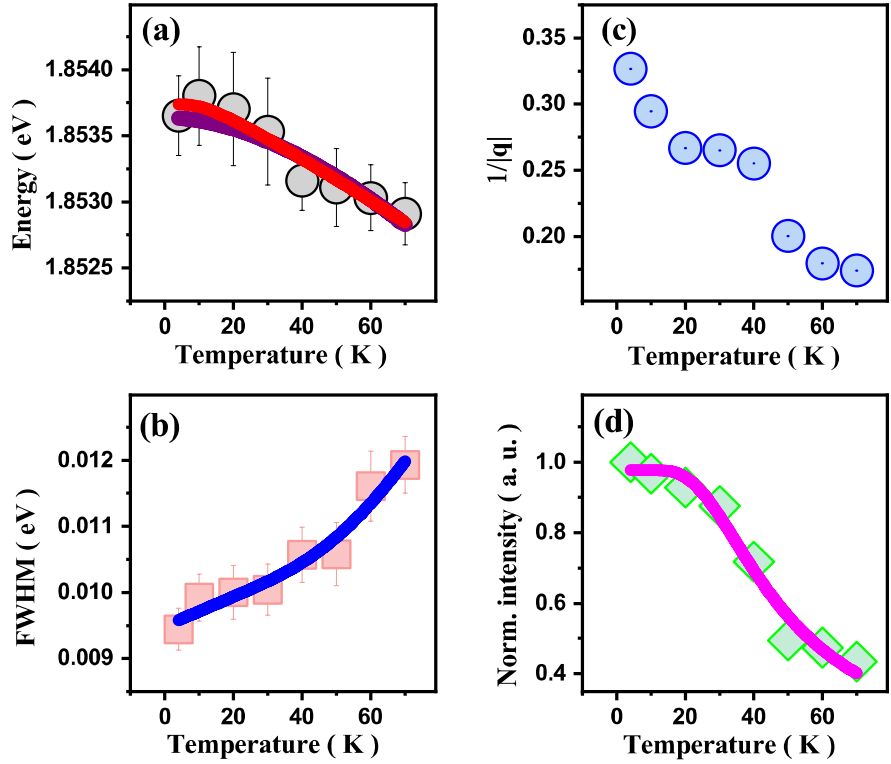

**Figure 2.** Temperature-dependent evolution of the excitonic peak (E1): **(a)** Peak energy (position), **(b)** Full width at half maximum (FWHM), **(c)** Fano asymmetry parameter (1/|q|), and **(d)** Integrated PL intensity. Solid red and purple lines in panel (a) represent model fits to the bandgap energy using Varshni and O'Donnell-Chen relations, respectively. Solid blue and magenta lines correspond to fitting of the FWHM and normalised intensity data in panels (b) and (d) as discussed in the text.

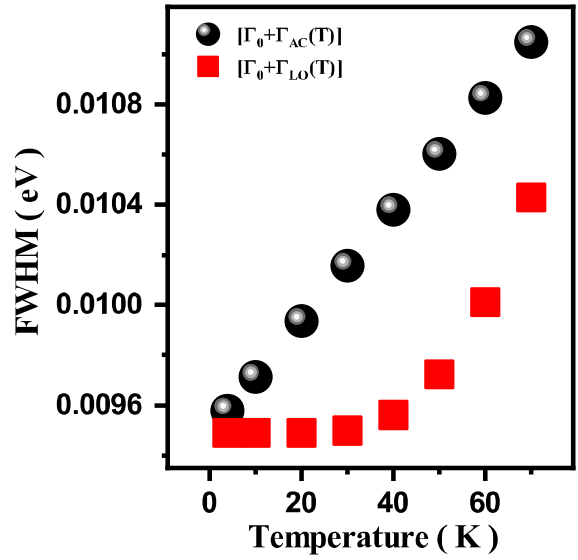

**Figure 3.** Individual contributions to the linewidth broadening of the excitonic peak E1 in bulk $Fe_2P_2S_6$, from acoustic and longitudinal optical (LO) phonons.

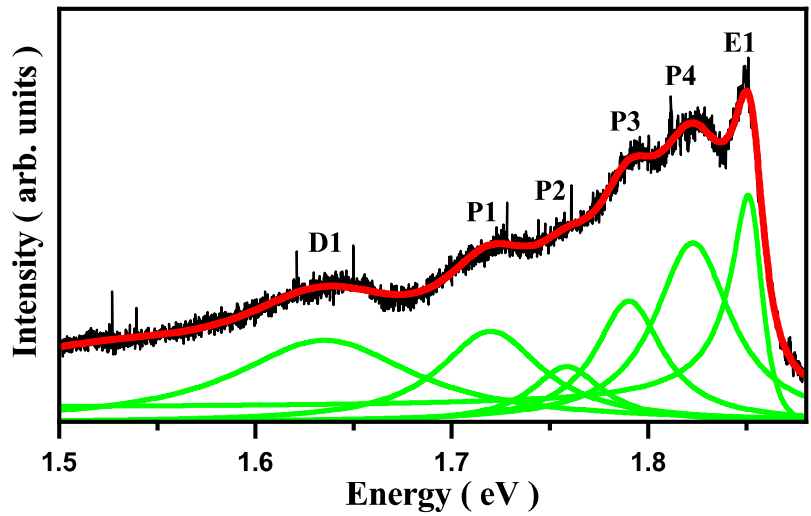

**Figure 4.** Fitted PL spectrum of bulk $Fe_2P_2S_6$ obtained at 4 K using excitation laser with wavelength 633 nm.

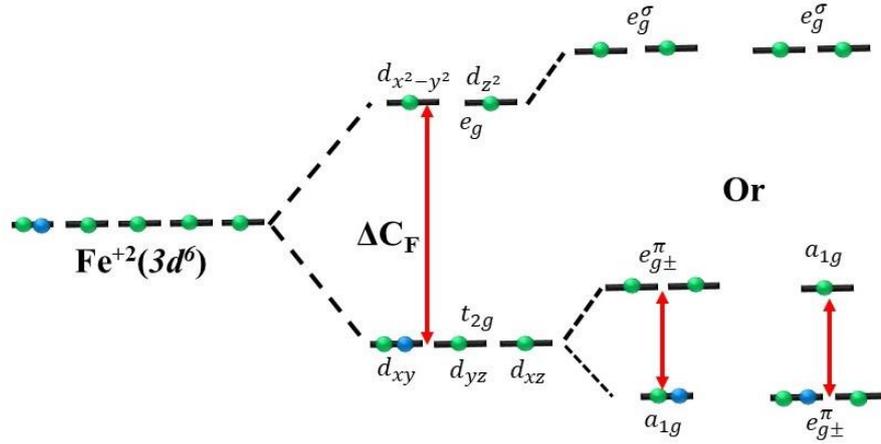

**Figure 5.** Schematic illustration of $Fe^{2+}$ $d$-orbital configurations. Left side depicts degenerate $3d$ orbitals in free-ion form; the central panel shows crystal field splitting ($\Delta C_F$) into $t_{2g}$ and $e_g$ levels under an octahedral ligand environment; and the right-side highlights further splitting of the $t_{2g}$ and $e_g$ manifolds induced by trigonal distortion.

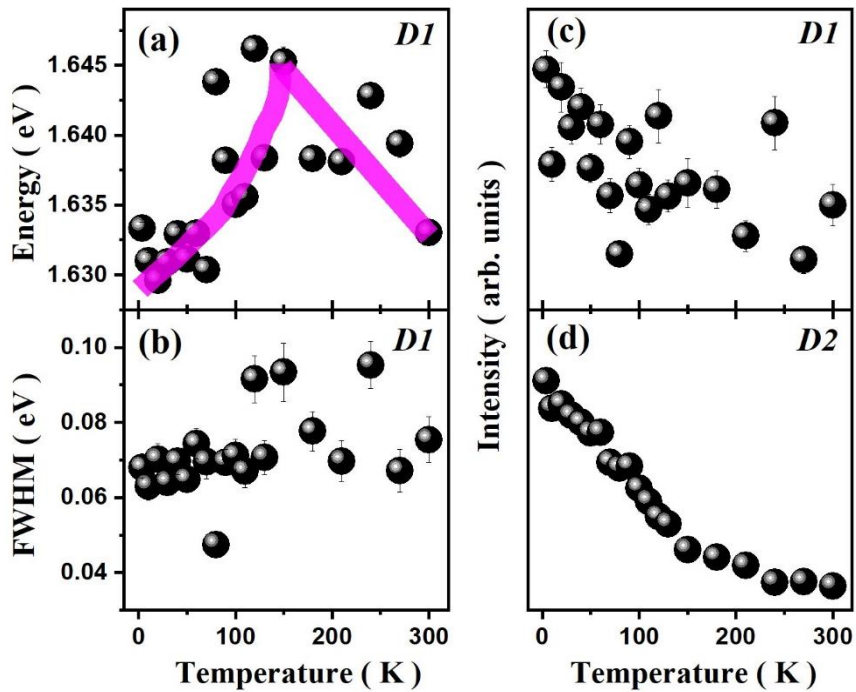

**Figure 6.** Temperature-dependent behaviour of $d$-$d$ transition peaks in bulk $Fe_2P_2S_6$. **(a)** Peak energy (position), **(b)** Full width at half maximum (FWHM), and **(c)** Integrated intensity for the spin-allowed $d$-$d$ transition (D1); **(d)** Temperature evolution of integrated intensity for the spin-forbidden $d$-$d$ transition (D2). Magenta solid line is guide to the eye.